\documentclass[12pt]{article}
\newcommand{\PRE}[1]{}       % Use if journal style
\usepackage{graphicx}

\newcommand{\eqref}[1]{Eq.~(\ref{#1})}

\newcommand{\be}{\begin{equation}}
\newcommand{\ee}{\end{equation}}

\newcommand{\bea}{\begin{eqnarray}}
\newcommand{\eea}{\end{eqnarray}}

\begin{document}

\begin{titlepage}
\noindent
\begin{flushright}
UH-511-1130-08  \\
UCI-2008-31 \\
\end{flushright}
\vspace{1cm}

\begin{center}
  \begin{Large}
    \begin{bf}
A Toy Model for Gauge-Mediation in Intersecting
Brane Models\\

    \end{bf}
  \end{Large}
\end{center}
\vspace{0.2cm}
\begin{center}
\begin{large}
Jason Kumar \\
\end{large}
  \vspace{0.3cm}
  \begin{it}
Department of Physics and Astronomy, University of Hawaii \\
        ~~Honolulu, HI  96822, USA \\
    and\\
Department of Physics, University of California, Irvine\\
    Irvine, CA  92697, USA\\
\vspace{0.1cm}
\end{it}
\end{center}

%\maketitle
\begin{abstract}
We discuss the phenomenology of a toy intersecting
brane model where
supersymmetry is dynamically broken in an open
string hidden sector and gauge-mediated to the
visible sector.  Scalar masses $\sim {\rm TeV}$ are
easily realizable, and R-symmetry is broken.
These ideas are easily generalizable to
other intersecting brane models.
\end{abstract}

\vspace{1cm}

%\begin{flushleft}
%hep-th/yymmnnn \\
August 2008
%\end{flushleft}

\end{titlepage}
%\pacs{PACS numbers: }
%]

\section{Introduction}

To utilize supersymmetry to understand the electroweak
scale, there are two broad questions which one must
answer.  Although supersymmetry stabilizes the scale of
electroweak symmetry breaking against quantum corrections,
it does not itself explain
why that scale is small in relation to an a priori more
natural scale such as the GUT (or Planck) scale.  Thus, there
has been much study of models where dynamics generates
supersymmetry-breaking at a scale
exponentially smaller than the natural scale of the
theory~\cite{DSB}.
In addition, it has long been known that supersymmetry cannot
be broken within the MSSM sector itself.  Instead, it must
be broken in some auxiliary sector, with its effects communicated
to MSSM sector via interactions.  There are several different
models for this type of mediation of supersymmetry-breaking
(for example, through gravitational interactions or through
gauge interactions), but each such mechanism has its advantages
and difficulties.

There have been recent attempts to tackle both of these
problems~\cite{Intriligator:2006dd,ISSfollowup,ISSgaugemed}.
String theory has a natural role to play in understanding
these questions.  In a low-energy effective field theory,
the parameters of the theory are simply given at the energy-scale
where the theory is defined; a real understanding of the
scale of those parameters depends on the microscopic physics
which has been integrated out.  If one has specific models
of a high-energy theory, then one can potentially find that
certain parameters are necessarily small, leading to a
small SUSY-breaking scales~\cite{Dine:2006gm}.  String
theoretic realizations of the Standard Model are a natural
starting point.

These classes of microscopic theories connect the question
of how SUSY is broken to the question of how SUSY-breaking
is mediated to the MSSM
sector.  For example, the two most natural ways of embedding the MSSM
within Type II string theory are by branes at singularities or by
intersecting branes.  In the former class gravity mediation occurs
rather generically, whereas in the latter class, gauge-mediated
supersymmetry-breaking (GMSB) can be more natural.  So a model
of dynamical supersymmetry breaking (DSB) which appears in one of
these broad classes is naturally tied
to a complementary mediation mechanism.
Moreover, these mediation mechanisms have very important
phenomenological consequences; for example,
GMSB can provide an elegant explanation for small
flavor-changing neutral currents.
There is thus great interest
in how this whole picture can be tied together in detail.
For example, Aharony, Kachru and Silverstein exhibited models of
``retrofitted'' DSB which naturally appear in the context of
branes at singularities~\cite{Aharony:2007db}.

On the other side,
in~\cite{Kumar:2007dw} it was shown that intersecting brane models
with non-trivial Fayet-Iliopoulos terms naturally generate
SUSY-breaking in the open-string sector at a scale exponentially
smaller than the FI-term scale.
In this note, we consider
an extension of the specific example of this class studied
in~\cite{Kumar:2007dw}, and examine the GMSB phenomenology
(note, Cvetic and Weigand
recently considered GMSB in a different IBM, which did
not utilize non-vanishing FI-terms~\cite{Cvetic:2008mh};
gauge-mediation in string models was also considered
in~\cite{Floratos:2006hs}).  In particular, we find that
scalar vevs break gauge symmetries at a relatively high
``natural" scale $\sqrt{\xi}$ set by the Fayet-Iliopoulos
terms.  But the SUSY-breaking $F$-terms are further
suppressed by exponentially small couplings in the superpotential,
$\langle F \rangle \sim \lambda \xi$.  The gauge messengers
then get masses $\sim \lambda \sqrt{\xi}$, 
yielding soft masses which are suppressed 
from the natural scale by
$m_{soft} \sim {g^2 \over 16\pi^2} \lambda \sqrt{\xi}$.

  In section 2 we will
describe the specific intersecting brane model and the details
of dynamical supersymmetry breaking.  In section 3 we will
study the phenomenology of gauge-mediation in this model, and
we will conclude with a discussion of further issues in section 4.

\section{Model}

For any string model to have realistic low-energy
phenomenology, it must somehow generate the chiral
matter content of the Standard Model.  In intersecting
brane models, the light chiral matter lies at the
intersection of D-branes which fill spacetime and
wrap cycles of the six compact
extra dimensions which are inherent to string theory.
In Type IIA string theory constructions, these are
D6-branes which fill spacetime and wrap a 3-cycle of
an orientifolded Calabi-Yau 3-fold.

Essentially, the gauge theory degrees of freedom arise
from strings which begin and end on the same stack of
D-branes, while chiral matter transforming in the
bifundamental representation of two gauge groups arises
from strings stretching between two brane stacks at their
point of intersection.  In particular, the number of
chiral multiplets transforming in the bifundamental of
groups $G_a$ and $G_b$ living on D-brane stacks $a$ and
$b$ is counted by $I_{ab}$, the topological intersection
number of the two brane
stacks\footnote{Chiral multiplets transforming in symmetric and
anti-symmetric representations will also arise from strings
stretching between a brane and its orientifold image, but
we will not need to deal with these multiplets here. }.

In \cite{Kumar:2007dw}, it was pointed out that intersecting
brane models can naturally exhibit dynamical supersymmetry breaking.
The key features here are that the Fayet-Iliopoulos terms are
non-zero and the Yukawa couplings $\lambda_i$
are exponentially small (because
they are generated by worldsheet instantons).  As a result, the
configuration space is constrained to live on $D$-flat directions
where the fields have vevs of order the FI-terms.  The $F$-terms
are then generically non-zero, and yield an $F$-term potential
$V_F \sim \lambda^2 \xi^2$ which is exponentially small compared
to the natural scale of the gauge theory set by the FI-terms.

The particular example described in \cite{Kumar:2007dw}
had three $U(1)$ gauge groups, and three chiral multiplets
which were each charged under two of the three gauge groups.
The three $U(1)$ FI-terms were taken to be non-zero, and the
three multiplets were coupled by a superpotential term with
an exponentially suppressed Yukawa coupling.

\begin{figure}[ht]
\centering
\includegraphics[width=12cm]{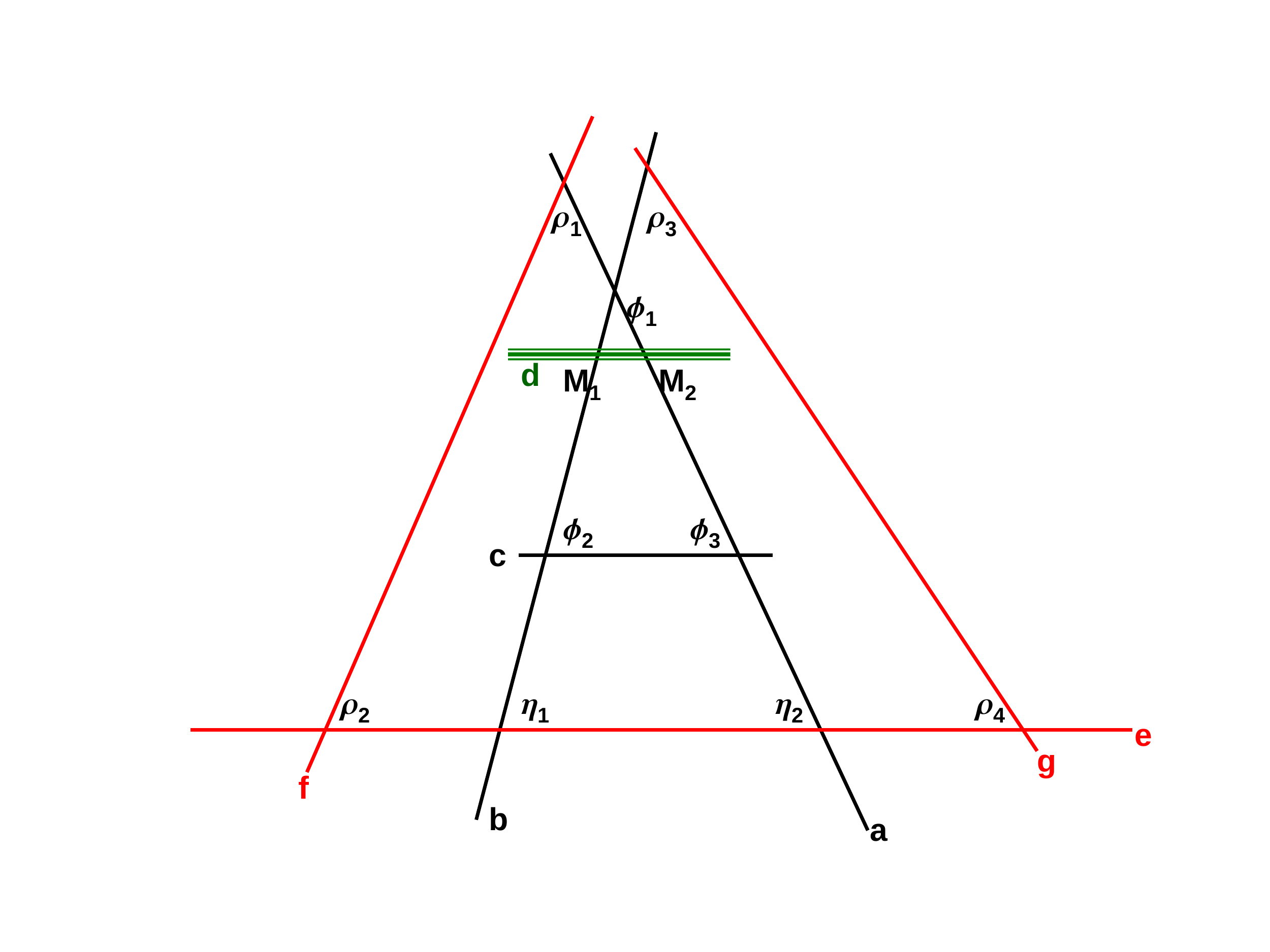}
\caption{This figure depicts the various brane stacks of this model
intersecting in the extra dimensions.  The brane stack $d$ is
a visible sector MSSM brane.  The brane stacks $a$, $b$ and $c$
are responsible for the relevant dynamical supersymmetry breaking.
the remaining stacks $e$, $f$ and $g$ are added to ensure that
cubic anomalies are canceled.  The multiplets
shown in the figure arise
from strings living at the indicated topological intersections.}
\label{setup}
\end{figure}

Here, we generalize the above model to include
the fields which mediate
supersymmetry breaking to a Standard Model sector
via gauge interactions.  The brane construction is depicted in
Fig.~\ref{setup}, and it is assumed
(for simplicity) that all brane stacks
generate $U(1)$ gauge groups.  The intersection numbers of the
various brane stacks are given by:
$I_{ba}=I_{ac}=I_{cb}=I_{ad}=I_{db}=I_{af}=I_{gb}=I_{fe} =
I_{eg}=1$; $I_{ea}=I_{be}=2$ (all others are zero).
The field content and
associated charges are summarized in
Table \ref{fieldcontent}.

In this setup, brane stacks $a$, $b$ and $c$ are the dynamical
SUSY-breaking sector described in \cite{Kumar:2007dw}.  The brane
stack $d$ is a Standard Model brane stack.  In an SU(5) GUT model,
it could be a brane stack where $U(5)_{GUT}$ lives.  Otherwise, it
could be a stack where $SU(2)_L$ lives.  The chiral multiplets
$M_{1,2}$ are the gauge messengers; although they are non-vectorlike
under hidden sector gauge groups, they are vectorlike under the
Standard Model.  The remaining branes are included to ensure that
the superpotential couplings are sufficiently generic and that
cubic anomalies are canceled (equivalently, that there are no
RR-tadpoles)\footnote{In general, more than one brane stack would
be required to construct a visible sector.  In this toy model the
additional branes
are not relevant to the SUSY-breaking phenomenology.
If the additional SM branes intersect the
branes of the DSB sector, then the superpotential will be more
complicated.  For the reasons described above
and in~\cite{Kumar:2007dw}, we still expect supersymmetry to be
dynamically broken, but obtaining reasonable phenomenology might
be more complicated.  We will assume no such
intersections here.}.

\begin{table}[t]
\caption{Field Content}
\centering
\begin{tabular}{l c c c c c c c}
%\hline
\, & $U(1)_a$ &  $U(1)_b$ &  $U(1)_c$ &  $U(1)_d$ &  $U(1)_e$
&  $U(1)_f$ &  $U(1)_g$ \\
\hline
$\phi_1$ & -1 & +1 & 0 & 0 & 0 & 0 & 0 \\
$\phi_2$ & 0 & -1 & +1 & 0 & 0 & 0 & 0 \\
$\phi_3$ & +1 & 0 & -1 & 0 & 0 & 0 & 0 \\
$M_1$ & 0 & -1 & 0 & +1 & 0 & 0 & 0 \\
$M_2$ & +1 & 0 & 0 & -1 & 0 & 0 & 0 \\
$\eta_1 ^{1,2}$ & 0 & +1 & 0 & 0 & -1 & 0 & 0 \\
$\eta_2 ^{1,2}$ & -1 & 0 & 0 & 0 & +1 & 0 & 0 \\
$\rho_1 $ & +1 & 0 & 0 & 0 & 0 & -1 & 0 \\
$\rho_2 $ & 0 & 0 & 0 & 0 & -1 & +1 & 0 \\
$\rho_3 $ & 0 & -1 & 0 & 0 & 0 & 0 & +1 \\
$\rho_4 $ & 0 & 0 & 0 & 0 & +1 & 0 & -1 \\
%\hline
\end{tabular}
\label{fieldcontent}
\end{table}

Without loss of generality, we may assume $|\eta_a|^2 =
|\eta_a ^1|^2 +|\eta_a ^2|^2$.
Given the charges of the various scalar fields, the general
$D$-term potential is given by
\bea
V_D &=&
{g_a ^2 \over 2} (|\phi_3|^2 - |\phi_1|^2 +|M_2 |^2
-|\eta_2|^2 +|\rho_1|^2-\xi_a)^2
\nonumber\\
&\,&+ {g_b ^2 \over 2} (|\phi_1|^2 - |\phi_2|^2 -|M_1 |^2
+|\eta_1|^2 -|\rho_3|^2-\xi_b)^2
\nonumber\\
&\,&+ {g_c ^2 \over 2} (|\phi_2|^2 - |\phi_3|^2 -\xi_c)^2
%\nonumber\\ &+&
+{g_d ^2 \over 2} (|M_1 |^2 -|M_2|^2 -\xi_d)^2
\nonumber\\
&\,&+ {g_e ^2 \over 2} (|\eta_2|^2 -|\eta_1|^2
+|\rho_4|^2 -|\rho_2|^2-\xi_e)^2
\nonumber\\
&\,&+ {g_f ^2 \over 2} (|\rho_2|^2 -|\rho_1|^2 -\xi_f)^2
%\nonumber\\ &+&
+{g_g ^2 \over 2} (|\rho_3|^2 -|\rho_4|^2 -\xi_g)^2
\eea
The Fayet-Iliopoulos terms $\xi_{a,...,g}$ can be non-zero,
and control the mass of the lightest scalar excitations
of the strings stretching between different brane stacks
(in the Type IIA picture, they are set by the angles between
the D6-brane stacks and the orientifold
planes~\cite{Berkooz:1996km}).  The
$\xi_{a...g}$ need not be identical, but in the interests
of simplicity of notation and of the model, we will assume
the non-zero FI-terms are all at about the same
approximate scale $\xi$.
The FI-terms are naturally set by the
string scale; however, if they are in fact at the string
scale, the effective field theory is not in a controlled
regime (since
the excited string excitations will be at the same scale as
light chiral modes).  To ensure that the effective field theory
approximation is valid, we will assume that the scale of the
FI-terms $\xi$ is slightly suppressed from the string scale;
the suppression need not be large.  Indeed, it would be
quite reasonable to set the $\xi$ at the GUT scale, if the
string scale is somewhere between the GUT and Planck
scales\footnote{In string theory, these FI-terms in fact
depend on closed-string moduli.  They may be treated as
constants only if (as we assume) those moduli are
stabilized
at a scale higher than the gauge-theory scale~\cite{Intriligator:2005aw}.}.

For the superpotential, we take
\be
W = \lambda_1 \phi_1 M_1 M_2 + \lambda_2 \phi_1 \phi_2 \phi_3
+\lambda_3 ^i \eta_1 ^{i} \rho_3 \rho_4
+\lambda_4 ^i \eta_2 ^i \rho_1 \rho_2 +\lambda_t \xi_t \phi_1 ,
\ee
where $i=1,2$. We will assume for simplicity
that $\lambda_{3} ^1= \lambda_{3} ^2$,
$\lambda_4 ^1 = \lambda_4 ^2$.
The first four terms constitute the most general renormalizable
superpotential which is classically gauge-invariant.
These Yukawa couplings are generated by Euclidean worldsheet
instantons which stretch between the three brane stacks that
the coupled fields begin and end
on~\cite{WSinstanton}.  Importantly, the
Yukawa couplings are exponentially suppressed by the area
of the instanton, $\lambda \propto e^{-{A/ l_s ^2}}$.
In particular, in the limit where the extra dimensions
are larger than the string scale (which is where most
mechanisms for moduli stabilization are effective), these
Yukawa couplings are expected to be exponentially
small\footnote{Note that in any intersecting brane model, it
is necessary for the top Yukawa coupling to be
${\cal O}(1)$.  This can arise from a worldsheet instanton
which is of string-scale size, which would require a modest
tuning if the compact dimensions are only moderately large. }.

In addition to these cubic terms,
there may be terms in the superpotential which naively are
not gauge-invariant.  These terms can arise because the $U(1)$
symmetries suffer from mixed anomalies, due to the presence of
matter transforming in non-vectorlike representations.  In
string theory, these mixed anomalies are fixed by the Green-Schwarz
mechanism; a closed-string axion field shifts under the
anomalous $U(1)$
and causes a classical transformation of the action which cancels
the anomalous loop contribution.  But if this axion field appears
in the exponent of a superpotential coupling, then it generates
a phase which can cancel the naively gauge-non-invariant phase
arising from the open string scalars.  This field is absorbed
into the coupling in the low-energy limit, below the scale
where closed-string moduli are stabilized.  The last term in the
superpotential is such a term, where $\xi_t$ is an energy scale
which is chosen to be of the same order as the FI-terms (we may do
this without loss of generality by rescaling the coupling; note
that the hierarchy between the scale $\xi$ and the string scale
is assumed to be relatively small).
Because this coupling is generated
by an E2-instanton, it is also
non-perturbatively small~\cite{strDinst,Cvetic:2008mh}.
The essential point about
the superpotential we have written is that, in terms of the
natural scale $\xi$, all terms in the superpotential have
dimensionless coefficients which are exponentially small
(and in particular, much smaller than the gauge couplings
$g_{a,...,g}$).  We will take all of
the $\lambda$'s to be real and positive.

For simplicity, we assume $\xi_d=0$.  Otherwise, we would be forced
to give a vev to the messengers and break SM or GUT symmetry at this
stage.  Although this might be desirable in certain GUT models,
we will avoid this complication\footnote{In a realistic model, $U(1)_Y$
should be FI-free.  This can arise if $U(1)_Y$ is a subgroup
of a GUT group (like $SU(5)$), or if $U(1)_Y$ arises as the FI-free
linear combination of various $U(1)$'s living on various D-branes}.
We have
$\sum_{i=a,...,g}\xi_i =0$, and we will assume that
$\xi_{e,g}<0$ and $\xi_{a,b,c}>0$ (this particular
choice will yield interesting
realistic phenomenology; there may be other choices which also
work).

Because the superpotential couplings are all exponentially suppressed,
the field configuration will sit close to a $D$-flat direction, but will
be shifted away from the $D$-flat direction by corrections which are
suppressed by ${\cal O}(\lambda_i ^2)$.
One solution to the $D$-term equations is given by
\bea
|\phi_1|^2 &=& \xi_b +\xi_c  \nonumber\\
|\phi_2|^2 &=& \xi_c \nonumber\\
|\rho_1|^2 &=& \xi_a +\xi_b +\xi_c \nonumber\\
|\rho_2|^2 &=& -\xi_e -\xi_g \nonumber\\
|\rho_4|^2 &=& -\xi_g \nonumber\\
\phi_3 =\eta_1 ^i =\eta_2 ^i =M_1 = M_2 =\rho_3 &=& 0
\eea
There are 5 $D$-flat directions away from this solution:
\bea
\delta |\phi_1 ^2| =\delta |\phi_2|^2=\delta |\phi_3|^2 &=& r_1,
\nonumber\\
\delta |\phi_1 ^2| =\delta |M_1|^2=\delta |M_2|^2 &=& r_2,
\nonumber\\
-\delta |\phi_1 ^2| =\delta |\eta_1|^2=\delta |\eta_2|^2 &=& r_3,
\nonumber\\
\delta |\rho_1 ^2| =\delta |\rho_2|^2=\delta |\eta_2|^2 &=& r_4,
\nonumber\\
\delta |\rho_3 ^2| =\delta |\rho_4|^2=\delta |\eta_1|^2 &=& r_5,
\eea
where we are constrained to $r_{1,...,5} \geq 0$.
The $\lambda_3$ terms in $V_F$
force the field configuration to $r_5 = 0$.  Similarly, the
$\lambda_4$ terms force us to
$r_4=0$.  However, $r_{1,2,3}$ are not yet constrained.
Plugging in the expansion above and keeping terms linear in
$r_{1,2,3}$ gives
\bea
V_F &=&
|\lambda_1 M_1 M_2 +\lambda_2 \phi_2 \phi_3 +\lambda_t \xi_t|^2
+\lambda_2 ^2 |\phi_1|^2 |\phi_2|^2 +\lambda_2^2 |\phi_1|^2 |\phi_3|^2
\nonumber\\
&\,&
+\lambda_1 ^2 |\phi_1|^2 |M_1|^2 +\lambda_1^2 |\phi_1|^2 |M_2|^2
+\lambda_3 ^2 |\eta_1|^2 |\rho_4|^2
+\lambda_3 ^2 |\eta_1|^2 |\rho_3|^2
\nonumber\\
&\,&
+\lambda_3 ^2 |\rho_3|^2 |\rho_4|^2
+\lambda_4 ^2 |\eta_2|^2 |\rho_1|^2
+\lambda_4 ^2 |\eta_2|^2 |\rho_2|^2
+ \lambda_4 ^2 |\rho_1|^2 |\rho_2|^2
\nonumber\\
&\sim&  \lambda_4 ^2 |(\xi_a +\xi_b +\xi_c)(\xi_e +\xi_g)|
+|\lambda_t \xi_t|^2
+\lambda_2 ^2 | \xi_c (\xi_b + \xi_c) |
\nonumber\\
&\,&
+\left[\lambda_2 ^2 (2\xi_b +4\xi_c)r_1
-2\lambda_2 \lambda_t |\xi_t| \sqrt{\xi_c} \sqrt{r_1}
\right]
\nonumber\\
&\,&
+\left[\lambda_2 ^2 \xi_c +2\lambda_1 ^2 (\xi_b + \xi_c)
-2\lambda_1 \lambda_t |\xi_t| \right]r_2
\nonumber\\
&\,&
+\left[\lambda_3 ^2 |\xi_g| +\lambda_4 ^2 (\xi_a +\xi_b +\xi_c
-\xi_e -\xi_g) -\lambda_2 ^2 \xi_c \right]r_3
\eea
The potential will force $r_{2,3}=0$ if the following
two constraints are satisfied:
\begin{itemize}
\item{$\lambda_2 ^2 \xi_c +2\lambda_1^2 (\xi_b +\xi_c)
>2\lambda_1 \lambda_t |\xi_t|$}
\item{$\lambda_3 ^2 |\xi_g| +\lambda_4 ^2
(\xi_a +\xi_b +\xi_c -\xi_e -\xi_g) > \lambda_2 ^2 \xi_c $}
\end{itemize}
Miminizing the potential with respect to $\sqrt{r_1}$ gives
\be
\sqrt{r_1} = {\lambda_t \over \lambda_2}
{\xi_t \sqrt{\xi_c} \over 2(\xi_b +2\xi_c)}.
\ee
If $\sqrt{r_1}$ is small, then the linear approximation
used above is valid.
Clearly, one can satisfy all three constraints (for example,
if all $\xi$ are of the same order and
$\lambda_t \ll \lambda_2 \ll \lambda_{3}$).
Supersymmetry is then broken, and the non-zero F-terms are
\bea
F_{\eta_2 ^i} &=& \lambda_4 ^i \sqrt{\xi_a +\xi_b +\xi_c}
\sqrt{|\xi_e +\xi_g|}\nonumber\\
F_{\phi_1} &=& \lambda_t \xi_t
\left( 1-{\xi_c \over 2(\xi_b +2\xi_c)}
\right)
\nonumber\\
F_{\phi_2} &=& \lambda_t  {\xi_t \sqrt{ \xi_c (\xi_b +\xi_c)}
\over 2(\xi_b +2\xi_c)}
\nonumber\\
F_{\phi_3} &=& \lambda_2 \sqrt{\xi_c} \sqrt{\xi_b +\xi_c}
+{\cal O} (\lambda_t  \xi ).
\eea
Non-zero $D$-terms arise because the field configuration is
shifted away from a $D$-flat direction by corrections due to
the $F$-term potential.
But the $D$-terms scale as
\be
V_D \sim {\cal O}\left({\lambda^4 \over g^2} \xi^2 \right)
\ee
and are much smaller than the $F$-terms in the limit where
the Yukawa couplings are exponentially
suppressed~\cite{Kumar:2007dw}.

\section{Phenomenology}

In the model we have described, the multiplets
$M_{1,2}$ naturally act as gauge messengers, which
are charged under both the MSSM sector and the
SUSY-breaking sector.  The mass-splitting between
the bosonic and fermionic components of the messenger
multiplets is generated by the superpotential term
\be
W = \lambda_1 \phi_1 M_1 M_2 +....
\ee
As such, the $F$-term relevant for GMSB phenomenology
in this model is $F_{\phi_1}$.

We can compute the soft scalar masses and find
\be
m_{scalar} \sim {g^2 \over 16\pi^2} {\lambda_1 F_{\phi_1}
\over \lambda_1 \langle \phi_1 \rangle} \sim
{g^2 \lambda_t  \over 16\pi^2} {\xi_t \over \sqrt{\xi_b +\xi_c}}
\left( 1-{\xi_c \over 2(\xi_b +2\xi_c)}
\right)
\ee

So if all $\xi$ are of the same natural scale, then the
soft scalars are at $m^2 \sim g^4 \lambda_t ^2 \xi$, and
are exponentially suppressed with respect to the scale $\xi$.
Furthermore, we have $W=\lambda_t \xi_t
\sqrt{\xi_b +\xi_c}$, so $R$-symmetry is broken and the
gauginos get a mass at the same scale as the scalars
(for this purpose, it was important that in our
solution $\langle \phi_1 \rangle , \langle F_{\phi_1} \rangle
\neq 0$).
The gravitino mass is given by $m_{3/2} \sim
{F \over M_{pl}}$, but in the limit we took earlier, the
dominant $F$-term is $F_{\phi_3} \sim \lambda_2 \xi$.
We would thus find
\bea
m_{3/2} &\sim & {F_{\phi_3} \over M_{Pl}}
\sim \lambda_2 {\xi \over M_{Pl}}.
\nonumber\\
{m_{3/2}^2 \over m_0 ^2}  &\sim &
{256 \pi^4 \lambda_2 ^2 \over g^4 \lambda_t ^2}
{\xi \over M_{Pl}^2} .
\eea

To get a feel for the phenomenology of a
somewhat natural example, let us assume that the scale
of the FI-terms is $\xi \sim (10^{15}\, {\rm GeV})^2$ ,
while the squark
and slepton masses are $\sim {\rm TeV}$.  Then from the above
formula we would find that $\lambda_t \sim 10^{-10}$.
This is not unreasonable if the extra dimensions
are (as is typical) stabilized at scales which are somewhat
larger than the string scale~\cite{modfix}, as
the coupling is generated by an E2-instanton~\cite{strDinst}
which is suppressed by $exp[-{2\pi \over
l_s ^3 g_s}Vol_{E2}]$.
Larger
Yukawa couplings could be accommodated by having
a smaller scale $\xi$.
Note that the messenger mass scale is
$m_{mess.} \sim \lambda_1 \xi$; since we have not yet
constrained the $\lambda_1$ coupling, messenger masses
from TeV to just below GUT scale are possible.

Assuming the point in parameter-space we described
above, we then find
\be
m_{3/2} \sim \lambda_2 \times 10^{11} {\rm GeV} .
\ee
Given our assumption $\lambda_2 > \lambda_t$, we
would find in this model a relatively heavy gravitino
with $m_{3/2} > 10\,{\rm GeV}$.  In particular,
the gravitinos here are typically heavier than those
found in \cite{Cvetic:2008mh}.
This seems quite reasonable; since the natural scale
of the theory is well above the intermediate scale,
even with suppressions of the $F$-terms from Yukawa
couplings, one would expect gravitational interactions
to be relatively large.  It might be even more natural
to have a slightly higher scale for $\xi$, say at
$10^{16}\,{\rm GeV}$.  However, in that case the
gravity-mediated contributions become significant
enough that they can potentially violate FCNC
constraints.  Since, absent an model-dependent explanation,
there is no reason for gravity contributions to respect
flavor, it is safest to allow a small hierarchy between
$\xi$ and the GUT scale.

In GMSB, gravitinos have a very interesting effect on
cosmology.  If R-symmetry is preserved, gravitinos are
stable and one must be sure that they do not overclose
the universe. For gravitinos in the mass-range discussed
here, the dominant production mechanism is through
sparticle scattering ~\cite{Giudice:1998bp},
and one finds that the gravitino density would be
phemenologically acceptable for reheat temperatures
$T_R < 10^{8-9} \,{\rm GeV}$ (or, alternatively, if
there were some mechanism of late entropy production).

It is worth noting also that in this model we have
introduced only one pair of messenger multiplets.
Although vectorlike matter transforming in complete
GUT group representations will not prevent gauge
coupling unification, they will change the scale at which
it occurs.  If enough matter multiplets are added,
the couplings may become strong before unification, making
the gauge-coupling unification unreliable.  One of the
advantages of SUSY-breaking sectors with Abelian dynamics
(such as the one described here) is that they typically
arise with small numbers of messengers, avoiding this
problem.

\section{Discussion}

We have exhibited a toy example of an intersecting brane
model in which supersymmetry is dynamically broken in
an open string hidden sector, and transmitted to the
visible SM sector via gauge interactions.  The natural
scale of the theory is set by the scale of the
Fayet-Iliopoulos terms, but the SUSY-breaking $F$-terms
are at
an exponentially smaller scale.  The soft masses in the
visible sector receive additional suppression from a loop
factor, and with reasonable choices of parameters the
gravitino is moderately lighter than scalars.  This hierarchy
of scales is summarized in Table \ref{scales}.

\begin{table}[t]
\caption{Scales}
\centering
\begin{tabular}{c l  }
%\hline
%\,  \\
\hline
%\hline
$\sqrt{\xi}$ & {\rm FI-terms and hidden gauge symmetry breaking}\\
$\lambda_1 \sqrt{\xi}$ & {\rm messenger mass scale}\\
$\sqrt{\lambda_1 \lambda_t} \sqrt{\xi}$ & {\rm messenger mass splitting} \\
${g^2 \over 16\pi^2} \lambda_t \sqrt{\xi}$ &
{\rm soft scalar masses}\\
$\lambda_2 {\xi\over M_{pl}}$ & {\rm gravitino mass}\\
%\hline
\end{tabular}
\label{scales}
\end{table}

It is interesting to note that this work illustrates
a point made in \cite{Kumar:2007dw}; if FI-terms
are turned on, one generically expects $F$-term
SUSY-breaking at exponentially small scales in
intersecting brane models.  Although
the model presented here extends the model in
\cite{Kumar:2007dw} by adding more fields and couplings,
the DSB result is robust.  Moreover, it was quite
easy to obtain realistic GMSB phenomenology.

There are a number of open questions worth studying.
From a formal standpoint, perhaps the most interesting
questions are related to moduli-stabilization.  In
the framework described above, we have essentially
treated the FI-terms as constants.  But in string
theory, the FI-terms depend on closed string moduli
(for example, in the Type IIA picture, the FI-terms
are set by the relative angles between the D6-branes,
and thus depend on the complex structure moduli).
For these parameters to be approximately constant,
they must be stabilized at a scale which is much
larger than the scale of the gauge theory. It has been
argued that this can be done in a controlled
manner~\cite{FIfix}, but
we have not utilized any particular moduli stabilization
scheme.  It would be very interesting to actually
construct a phenomenologically viable model of gauge-mediated
dynamical supersymmetry breaking where these issues of
moduli-stabilization are concretely addressed.

In is interesting to note that the relative hierarchies
at phenomenological scales (say, between $m_{3/2}$ and
$m_0$) are controlled by different Yukawa couplings.
As a result, large mass hierarchies are obtainable from
relatively small hierarchies in areas in the extra
dimensions.  This implies that detailed phenomenology
in these types of models can depend crucially on how
moduli and brane positions are stabilized in the extra
dimensions.  It would be very interesting to understand
how these issues can be addressed concretely in
orientifolded Calabi-Yau 3-folds compactifications.

\section{Acknowledgments} This work was supported by NSF grants
PHY--0239817, PHY--0314712 and PHY--0653656.
We are grateful to J. Feng,
S. Franco, A. Rajaraman, B. Thomas and J. Wells
for useful discussions, and
to S. Kachru and E. Silverstein for collaboration at an early
stage of this project.

%%%%%%%%%%%%%%%%%%%%%%%%%%%%%%%%%%%%%%%%%%%%%%%%%%%%%%

%%%%%%%%%%%%%%%%%%%%%%%%%%%%%%%%%%%%%%%%%%%%%%%%%%%%%%


\begin{thebibliography}{99}
%%%%%%%%%%%%%%%%%%%%%%%%%%%%%%%%%%%%%%%%%%%%%%%%%%%%%%

\bibitem{DSB}
For reviews, see
%\cite{Affleck:1984xz}
%\bibitem{Affleck:1984xz}
  I.~Affleck, M.~Dine and N.~Seiberg,
  %``Dynamical Supersymmetry Breaking In Four-Dimensions And Its
  %Phenomenological Implications,''
  Nucl.\ Phys.\  B {\bf 256}, 557 (1985);
  %%CITATION = NUPHA,B256,557;%%
%\cite{Poppitz:1998vd}
%\bibitem{Poppitz:1998vd}
  E.~Poppitz and S.~P.~Trivedi,
  %``Dynamical supersymmetry breaking,''
  Ann.\ Rev.\ Nucl.\ Part.\ Sci.\  {\bf 48}, 307 (1998)
  [arXiv:hep-th/9803107];
  %%CITATION = ARNUA,48,307;%%
  %\cite{Shadmi:1999jy}
%\bibitem{Shadmi:1999jy}
  Y.~Shadmi and Y.~Shirman,
  %``Dynamical supersymmetry breaking,''
  Rev.\ Mod.\ Phys.\  {\bf 72}, 25 (2000)
  [arXiv:hep-th/9907225];
  %%CITATION = RMPHA,72,25;%%
%\cite{Intriligator:2007cp}
%\bibitem{Intriligator:2007cp}
  K.~Intriligator and N.~Seiberg,
  %``Lectures on Supersymmetry Breaking,''
  arXiv:hep-ph/0702069.
  %%CITATION = HEP-PH/0702069;%%
See also
%\cite{Cvetic:2003yd}
%\bibitem{Cvetic:2003yd}
  M.~Cvetic, P.~Langacker and J.~Wang,
  %``Dynamical supersymmetry breaking in standard-like models with  intersecting
  %D6-branes,''
  Phys.\ Rev.\  D {\bf 68}, 046002 (2003)
  [arXiv:hep-th/0303208].
  %%CITATION = PHRVA,D68,046002;%%

%\cite{Intriligator:2006dd}
\bibitem{Intriligator:2006dd}
  K.~Intriligator, N.~Seiberg and D.~Shih,
  %``Dynamical SUSY breaking in meta-stable vacua,''
  JHEP {\bf 0604}, 021 (2006)
  [arXiv:hep-th/0602239].
  %%CITATION = JHEPA,0604,021;%%

\bibitem{ISSfollowup}
For example, see
%\cite{Ooguri:2006bg}
%\bibitem{Ooguri:2006bg}
  H.~Ooguri and Y.~Ookouchi,
  %``Meta-stable supersymmetry breaking vacua on intersecting branes,''
  Phys.\ Lett.\  B {\bf 641}, 323 (2006)
  [arXiv:hep-th/0607183];
  %%CITATION = PHLTA,B641,323;%%
%\cite{Argurio:2007qk}
%\bibitem{Argurio:2007qk}
  R.~Argurio, M.~Bertolini, S.~Franco and S.~Kachru,
  %``Metastable vacua and D-branes at the conifold,''
  JHEP {\bf 0706}, 017 (2007)
  [arXiv:hep-th/0703236];
  %%CITATION = JHEPA,0706,017;%%
%\cite{Intriligator:2007py}
%\bibitem{Intriligator:2007py}
  K.~Intriligator, N.~Seiberg and D.~Shih,
  %``Supersymmetry Breaking, R-Symmetry Breaking and Metastable Vacua,''
  JHEP {\bf 0707}, 017 (2007)
  [arXiv:hep-th/0703281];
  %%CITATION = JHEPA,0707,017;%%
%\cite{Aganagic:2007py}
%\bibitem{Aganagic:2007py}
  M.~Aganagic, C.~Beem and S.~Kachru,
  %``Geometric Transitions and Dynamical SUSY Breaking,''
  Nucl.\ Phys.\  B {\bf 796}, 1 (2008)
  [arXiv:0709.4277 [hep-th]];
  %%CITATION = NUPHA,B796,1;%%
  %\cite{Marsano:2008ts}
%\bibitem{Marsano:2008ts}
  J.~Marsano, K.~Papadodimas and M.~Shigemori,
  %``Off-shell M5 Brane, Perturbed Seiberg-Witten Theory, and Metastable
  %Vacua,''
  arXiv:0801.2154 [hep-th];
  %%CITATION = ARXIV:0801.2154;%%
%\cite{Beasley:2008kw}
%\bibitem{Beasley:2008kw}
  C.~Beasley, J.~J.~Heckman and C.~Vafa,
  %``GUTs and Exceptional Branes in F-theory - II: Experimental Predictions,''
  arXiv:0806.0102 [hep-th];
  %%CITATION = ARXIV:0806.0102;%%
  %\cite{Dienes:2008gj}
%\bibitem{Dienes:2008gj}
  K.~R.~Dienes and B.~Thomas,
  %``Building a Nest at Tree Level: Classical Metastability and Non-Trivial
  %Vacuum Structure in Supersymmetric Field Theories,''
  arXiv:0806.3364 [hep-th];
  %%CITATION = ARXIV:0806.3364;%%
%\cite{Heckman:2008es}
%\bibitem{Heckman:2008es}
  J.~J.~Heckman, J.~Marsano, N.~Saulina, S.~Schafer-Nameki and C.~Vafa,
  %``Instantons and SUSY breaking in F-theory,''
  arXiv:0808.1286 [hep-th].
  %%CITATION = ARXIV:0808.1286;%%




\bibitem{ISSgaugemed}
%\cite{Kitano:2006xg}
%\bibitem{Kitano:2006xg}
  R.~Kitano, H.~Ooguri and Y.~Ookouchi,
  %``Direct mediation of meta-stable supersymmetry breaking,''
  Phys.\ Rev.\  D {\bf 75}, 045022 (2007)
  [arXiv:hep-ph/0612139];
  %%CITATION = PHRVA,D75,045022;%%
%\cite{Murayama:2006yf}
%\bibitem{Murayama:2006yf}
  H.~Murayama and Y.~Nomura,
  %``Gauge mediation simplified,''
  Phys.\ Rev.\ Lett.\  {\bf 98}, 151803 (2007)
  [arXiv:hep-ph/0612186];
  %%CITATION = PRLTA,98,151803;%%
%\cite{Murayama:2007fe}
%\bibitem{Murayama:2007fe}
  H.~Murayama and Y.~Nomura,
  %``Simple scheme for gauge mediation,''
  Phys.\ Rev.\  D {\bf 75}, 095011 (2007)
  [arXiv:hep-ph/0701231];
  %%CITATION = PHRVA,D75,095011;%%
%\cite{Kawano:2007ru}
%\bibitem{Kawano:2007ru}
  T.~Kawano, H.~Ooguri and Y.~Ookouchi,
  %``Gauge Mediation in String Theory,''
  Phys.\ Lett.\  B {\bf 652}, 40 (2007)
  [arXiv:0704.1085 [hep-th]];
  %%CITATION = PHLTA,B652,40;%%
%\cite{Haba:2007rj}
%\bibitem{Haba:2007rj}
  N.~Haba and N.~Maru,
  %``A Simple Model of Direct Gauge Mediation of Metastable Supersymmetry
  %Breaking,''
  Phys.\ Rev.\  D {\bf 76}, 115019 (2007)
  [arXiv:0709.2945 [hep-ph]];
  %%CITATION = PHRVA,D76,115019;%%
%\cite{Buican:2008qe}
%\bibitem{Buican:2008qe}
  M.~Buican and S.~Franco,
  %``SUSY breaking mediation by D-brane instantons,''
  arXiv:0806.1964 [hep-th].
  %%CITATION = ARXIV:0806.1964;%%



%\cite{Dine:2006gm}
\bibitem{Dine:2006gm}
  M.~Dine, J.~L.~Feng and E.~Silverstein,
  %``Retrofitting O'Raifeartaigh models with dynamical scales,''
  Phys.\ Rev.\  D {\bf 74}, 095012 (2006)
  [arXiv:hep-th/0608159].
  %%CITATION = PHRVA,D74,095012;%%

%\cite{Aharony:2007db}
\bibitem{Aharony:2007db}
  O.~Aharony, S.~Kachru and E.~Silverstein,
  %``Simple Stringy Dynamical SUSY Breaking,''
  Phys.\ Rev.\  D {\bf 76}, 126009 (2007)
  [arXiv:0708.0493 [hep-th]].
  %%CITATION = PHRVA,D76,126009;%%

%\cite{Kumar:2007dw}
\bibitem{Kumar:2007dw}
  J.~Kumar,
  %``Dynamical SUSY Breaking in Intersecting Brane Models,''
  Phys.\ Rev.\  D {\bf 77}, 046010 (2008)
  [arXiv:0708.4116 [hep-th]].
  %%CITATION = PHRVA,D77,046010;%%

%\cite{Cvetic:2008mh}
\bibitem{Cvetic:2008mh}
  M.~Cvetic and T.~Weigand,
  %``A string theoretic model of gauge mediated supersymmetry beaking,''
  arXiv:0807.3953 [hep-th].
  %%CITATION = ARXIV:0807.3953;%%

%\cite{Floratos:2006hs}
\bibitem{Floratos:2006hs}
  E.~Floratos and C.~Kokorelis,
  %``MSSM GUT string vacua, split supersymmetry and fluxes,''
  arXiv:hep-th/0607217.
  %%CITATION = HEP-TH/0607217;%%

%\cite{Berkooz:1996km}
\bibitem{Berkooz:1996km}
  M.~Berkooz, M.~R.~Douglas and R.~G.~Leigh,
  %``Branes intersecting at angles,''
  Nucl.\ Phys.\  B {\bf 480}, 265 (1996)
  [arXiv:hep-th/9606139];
  %%CITATION = NUPHA,B480,265;%%
  %\cite{Ohta:1997fr}
%\bibitem{Ohta:1997fr}
  N.~Ohta and P.~K.~Townsend,
  %``Supersymmetry of M-branes at angles,''
  Phys.\ Lett.\  B {\bf 418}, 77 (1998)
  [arXiv:hep-th/9710129].
  %%CITATION = PHLTA,B418,77;%%



%\cite{Intriligator:2005aw}
\bibitem{Intriligator:2005aw}
  K.~Intriligator and N.~Seiberg,
  %``The runaway quiver,''
  JHEP {\bf 0602}, 031 (2006)
  [arXiv:hep-th/0512347].
  %%CITATION = JHEPA,0602,031;%%

\bibitem{WSinstanton}
%\cite{Aldazabal:2000cn}
%\bibitem{Aldazabal:2000cn}
  G.~Aldazabal, S.~Franco, L.~E.~Ibanez, R.~Rabadan and A.~M.~Uranga,
  %``Intersecting brane worlds,''
  JHEP {\bf 0102}, 047 (2001)
  [arXiv:hep-ph/0011132];
  %%CITATION = JHEPA,0102,047;%%
%\cite{Kachru:2000ih}
%\bibitem{Kachru:2000ih}
  S.~Kachru, S.~H.~Katz, A.~E.~Lawrence and J.~McGreevy,
  %``Open string instantons and superpotentials,''
  Phys.\ Rev.\  D {\bf 62}, 026001 (2000)
  [arXiv:hep-th/9912151];
  %%CITATION = PHRVA,D62,026001;%%
%\cite{Cremades:2003qj}
%\bibitem{Cremades:2003qj}
  D.~Cremades, L.~E.~Ibanez and F.~Marchesano,
  %``Yukawa couplings in intersecting D-brane models,''
  JHEP {\bf 0307}, 038 (2003)
  [arXiv:hep-th/0302105];
%%CITATION = JHEPA,0307,038;%%
 M.~Cvetic and I.~Papadimitriou,
% ``Conformal field theory couplings for intersecting D-branes on
% orientifolds,''
  Phys.\ Rev.\  D {\bf 68}, 046001 (2003)
  [Erratum-ibid.\  D {\bf 70}, 029903 (2004)]
  [arXiv:hep-th/0303083].

\bibitem{strDinst}
%\cite{Ganor:1996pe}
%\bibitem{Ganor:1996pe}
  O.~J.~Ganor,
  %``A note on zeroes of superpotentials in F-theory,''
  Nucl.\ Phys.\  B {\bf 499}, 55 (1997)
  [arXiv:hep-th/9612077];
  %%CITATION = NUPHA,B499,55;%%
 R.~Blumenhagen, M.~Cvetic and T.~Weigand,
%``Spacetime instanton corrections in 4D string vacua - the seesaw
%mechanism for D-brane models,''
  Nucl.\ Phys.\  B {\bf 771}, 113 (2007)
  [arXiv:hep-th/0609191];
L.~E.~Ibanez and A.~M.~Uranga,
% ``Neutrino Majorana masses from string theory instanton effects,''
  JHEP {\bf 0703}, 052 (2007)
  [arXiv:hep-th/0609213];
%\cite{Florea:2006si}
%\bibitem{Florea:2006si}
  B.~Florea, S.~Kachru, J.~McGreevy and N.~Saulina,
  %``Stringy instantons and quiver gauge theories,''
  JHEP {\bf 0705}, 024 (2007)
  [arXiv:hep-th/0610003];
  %%CITATION = JHEPA,0705,024;%%
%\cite{Cvetic:2008ws}
%\bibitem{Cvetic:2008ws}
  M.~Cvetic, R.~Richter and T.~Weigand,
  %``(Non-)BPS bound states and D-brane instantons,''
  JHEP {\bf 0807}, 012 (2008)
  [arXiv:0803.2513 [hep-th]].
  %%CITATION = JHEPA,0807,012;%%



\bibitem{modfix}
For reviews, see
%\cite{Silverstein:2004id}
%\bibitem{Silverstein:2004id}
  E.~Silverstein,
  %``TASI / PiTP / ISS lectures on moduli and microphysics,''
  arXiv:hep-th/0405068;
  %%CITATION = HEP-TH/0405068;%%
%\cite{Grana:2005jc}
%\bibitem{Grana:2005jc}
  M.~Grana,
  %``Flux compactifications in string theory: A comprehensive review,''
  Phys.\ Rept.\  {\bf 423}, 91 (2006)
  [arXiv:hep-th/0509003];
  %%CITATION = PRPLC,423,91;%%
%\cite{Douglas:2006es}
%\bibitem{Douglas:2006es}
  M.~R.~Douglas and S.~Kachru,
  %``Flux compactification,''
  Rev.\ Mod.\ Phys.\  {\bf 79}, 733 (2007)
  [arXiv:hep-th/0610102];
  %%CITATION = RMPHA,79,733;%%
%\cite{Blumenhagen:2006ci}
%\bibitem{Blumenhagen:2006ci}
  R.~Blumenhagen, B.~Kors, D.~Lust and S.~Stieberger,
  %``Four-dimensional String Compactifications with D-Branes, Orientifolds   and
  %Fluxes,''
  Phys.\ Rept.\  {\bf 445}, 1 (2007)
  [arXiv:hep-th/0610327];
  %%CITATION = PRPLC,445,1;%%
%\cite{Denef:2007pq}
%\bibitem{Denef:2007pq}
  F.~Denef, M.~R.~Douglas and S.~Kachru,
  %``Physics of string flux compactifications,''
  Ann.\ Rev.\ Nucl.\ Part.\ Sci.\  {\bf 57}, 119 (2007)
  [arXiv:hep-th/0701050].
  %%CITATION = ARNUA,57,119;%%




\bibitem{FIfix}
%\cite{Binetruy:2004hh}
%\bibitem{Binetruy:2004hh}
  P.~Binetruy, G.~Dvali, R.~Kallosh and A.~Van Proeyen,
  %``Fayet-Iliopoulos terms in supergravity and cosmology,''
  Class.\ Quant.\ Grav.\  {\bf 21}, 3137 (2004)
  [arXiv:hep-th/0402046];
  %%CITATION = CQGRD,21,3137;%%
%\cite{Diaconescu:2007ah}
%\bibitem{Diaconescu:2007ah}
  D.~E.~Diaconescu, R.~Donagi and B.~Florea,
  %``Metastable quivers in string compactifications,''
  Nucl.\ Phys.\  B {\bf 774}, 102 (2007)
  [arXiv:hep-th/0701104];
  %%CITATION = NUPHA,B774,102;%%
%\cite{Sinha:2007rg}
%\bibitem{Sinha:2007rg}
  K.~Sinha,
  %``Stabilizing the Runaway Quiver in Supergravity,''
  arXiv:0709.2932 [hep-th].
  %%CITATION = ARXIV:0709.2932;%%



%\cite{Giudice:1998bp}
\bibitem{Giudice:1998bp}
  G.~F.~Giudice and R.~Rattazzi,
  %``Theories with gauge-mediated supersymmetry breaking,''
  Phys.\ Rept.\  {\bf 322}, 419 (1999)
  [arXiv:hep-ph/9801271];
  %%CITATION = PRPLC,322,419;%%
%\cite{Feng:2005ee}
%\bibitem{Feng:2005ee}
  J.~L.~Feng,
  %``Supersymmetry and cosmology,''
  Annals Phys.\  {\bf 315}, 2 (2005).
  %%CITATION = APNYA,315,2;%%


%%%%%%%%%%%%%%%%%%%%%%%%%%%%%%%%%%%%%%%%%%%%%%%%%%%%%%
\end{thebibliography}
\end{document}